\documentclass[10pt,a4papper]{article}
\usepackage{indentfirst}
\usepackage[]{graphicx}
\usepackage{geometry} 
\begin{document}

\title{\textbf{The decay $\tau \rightarrow K^{-}\pi^{0}\nu_{\tau}$ in the Nambu-Jona-Lasinio model}}
\author{M. K. Volkov\footnote{volkov@theor.jinr.ru}, A. A. Pivovarov\footnote{tex$\_$k@mail.ru}\\
\small
\emph{Bogoliubov Laboratory of Theoretical Physics, JINR, Dubna, 141980, Russia}}
\maketitle
\small

\begin{abstract}
In this paper, the width of the decay $\tau \rightarrow K^{-}\pi^{0}\nu_{\tau}$ is calculated
in the framework of the Nambu-Jona-Lasinio model. The contributions of the intermediate vector $K^{*}(892)$
and scalar $K_{0}^{*}(800)$ mesons are taken into account. It is shown that the main contribution to the width
of this decay is given by the subprocesses with the intermediate $W$-boson and vector $K^{*}(892)$ meson.
The scalar channel with the intermediate $K_{0}^{*}(800)$ meson gives an insignificant contribution. In Appendix,
it is shown that the contribution of the subprocess with the intermediate ${K^{*}}'(1410)$ meson is negligible as well.
The obtained results are in satisfactory agreement with the experimental data.
\end{abstract}
\large
\section{Introduction}
Experimental research of the $\tau$-lepton decays is intensively carried out at different scientific centers,
such as BaBar, Belle, etc. At an energy bellow 1.8 GeV ($m_{\tau} = 1.777$ GeV) the well-known perturbation theory
of quantum chromodynamics (QCD) cannot be applied for these processes due to a large value of the coupling constant. Thus, instead of the fundamental QCD
one has to use various phenomenological models. They are generally based on the chiral symmetry and use the
vector meson dominance model \cite{LopezCastro:1996xh, Finkemeier:1996dh, Li:1996md, Andrianov:2005kx, Jamin:2006tk,
Jamin:2008qg, Boito:2008fq, Boito:2010me, Dumm:2012vb, Calderon:2012zi, Escribano:2014joa, Kang:2013jaa}.
However, most of these models include a number of fitting parameters, thus reducing their predictive power.
The Nambu-Jona-Lasinio (NJL) model used in the present work does not have such defects and allows
describing experimental data without additional arbitrary parameters. In recent years, these qualities of the NJL model
were illustrated for example by successful description of such $\tau$-lepton decays as
$\tau \rightarrow \pi(\pi'(1300)) \nu_{\tau}$ \cite{Ahmadov:2015zua}, $\tau \rightarrow \rho(770)(\rho'(1450)) \nu_{\tau}$ \cite{Ahmadov:2015oca},
$\tau \rightarrow \pi^{-} \pi^{0} \nu_{\tau}$ \cite{Volkov:2012uh},
$\tau \rightarrow \eta(\eta') \pi \nu_{\tau}$ \cite{Volkov:2012be}, $\tau \rightarrow \pi \omega \nu_{\tau}$ \cite{Volkov:2012gv},
$\tau \rightarrow f_{1} \pi \nu_{\tau}$ \cite{Vishneva:2014lla}, $\tau \rightarrow \eta(\eta') 2\pi \nu_{\tau}$ \cite{Volkov:2013zba}.

Until this year, we had not considered the $\tau$-lepton decays with strange particles in final state.
However, we hope that this model allows describing these processes. Indeed, there was recently successfully
calculated the decay $\tau \rightarrow K^{*}(892)({K^{*}}'(1410)) \nu_{\tau}$ using the NJL model. In the present paper,
we continue to research in this direction and calculate the decay width of the process $\tau \rightarrow K^{-}\pi^{0}\nu_{\tau}$
in the framework of the standard NJL model. This model is intended for describing the four meson nonets only in the ground states
\cite{Volkov:1984kq, Volkov:1986zb, Ebert:1985kz, Vogl:1991qt, Klevansky:1992qe, Volkov:1993jw, Ebert:1994mf, Volkov:2001ns, Volkov:2006vq}.
The estimation of contributions from both vector and scalar channels of this process is given. The obtained results are
in satisfactory agreement with the experimental data. In Appendix, it is shown that the contribution of the subprocess
with the intermediate radially excited vector ${K^{*}}'(1410)$ meson is negligible as well as the contribution from diagram with scalar meson.
The extended NJL model \cite{Volkov:2006vq, Volkov:1996br, Volkov:1996fk, Volkov:1997dd, Volkov:1999yi} was used for estimation of this contribution.

\section{The Lagrangian of the standard NJL model and the amplitudes of the process $\tau \rightarrow K^{-}\pi^{0}\nu_{\tau}$}
In the standard NJL model, the quark-meson interaction Lagrangian for pseudoscalar $K^{\pm}, \pi^{0}$, scalar $K_{0}^{*\pm}$ and
vector $K^{*\pm}$ mesons takes the form:
\begin{displaymath}
\Delta L_{int}(q,\bar{q},\pi^{0},K,K_{0}^{*},K^{*}) = \bar{q}\left[ig_{\pi}\gamma^{5}\lambda_{3}\pi^{0} +
ig_{K}\gamma^{5}\lambda_{\pm}K^{\pm} \right.
\end{displaymath}
\begin{equation}
\left. + g_{K_{0}^{*}}\lambda_{\pm}K_{0}^{*\pm} +
\frac{g_{K^{*}}}{2}\gamma^{\mu}\lambda_{\pm}K^{*\pm}_{\mu}\right]q,
\end{equation}
where $q$ and $\bar{q}$ are the u-, d- and s- constituent quark fields with masses $m_{u} = m_{d} = 280$ MeV,
$m_{s} = 420$ MeV \cite{Volkov:2001ns}, $\pi^{0}$, $K^{\pm}$, $K^{*\pm}_{0}$ and $K^{*\pm}$ are the pseudoscalar, scalar and vector mesons, and
\begin{equation}
\lambda_{+} = \frac{\lambda_{4} + i\lambda_{5}}{\sqrt{2}} = \sqrt{2} \left(\begin{array}{ccc}
0 & 0 & 1\\
0 & 0 & 0\\
0 & 0 & 0
\end{array}\right), \quad
\lambda_{-} = \frac{\lambda_{4} - i\lambda_{5}}{\sqrt{2}} = \sqrt{2} \left(\begin{array}{ccc}
0 & 0 & 0\\
0 & 0 & 0\\
1 & 0 & 0
\end{array}\right),
\end{equation}
$\lambda_{3}, \lambda_{4}$ and $\lambda_{5}$ are the Gell-Mann matrices.

The coupling constants:
\begin{displaymath}
g_{\pi} = \sqrt{Z_{\pi}}g(m_{u}, m_{u}), \quad g_{K} = \sqrt{Z_{K}}g_{K_{0}^{*}},
\end{displaymath}
\begin{equation}
\label{Constants}
g_{K_{0}^{*}} = g(m_{u}, m_{s}), \quad g_{K^{*}} = \sqrt{6}g_{K_{0}^{*}},
\end{equation}
where
\begin{displaymath}
g(m_{1}, m_{2}) = (4I_{2}(m_{1}, m_{2}))^{-1/2},
\end{displaymath}
\begin{equation}
Z_{\pi} = \left(1 - 6\frac{m^{2}_{u}}{M^{2}_{a_{1}}}\right)^{-1}, \quad
Z_{K} = \left(1 - \frac{3}{2}\frac{(m_{u} + m_{s})^{2}}{M^{2}_{K_{1}}}\right)^{-1},
\end{equation}
$Z_{\pi}$ is the factor corresponding to the $\pi - a_{1}$ transitions,
$Z_{K}$ is the factor corresponding to the $K - K_{1}$ transitions,
$M_{a_{1}} = 1230$ MeV, $M_{K_{1}} = 1270$ MeV \cite{Agashe:2014kda} are the masses of the axial-vector
$a_{1}$ and $K_{1}$ mesons, and the integral $I_{2}$ in Euclidean space has the following form:
\begin{equation}
I_{2}(m_{1}, m_{2}) = \frac{N_{c}}{(2\pi)^{4}}\int\frac{\theta(\Lambda_{4}^{2} - k^2)}{(m_{1}^{2} + k^2)(m_{2}^{2} + k^2)}
\mathrm{d}^{4}k,
\end{equation}
$\Lambda_{4} = 1.25$ GeV is the cut-off parameter \cite{Volkov:1986zb}.

\begin{figure}[b]
\centerline{\includegraphics[scale = 0.7]{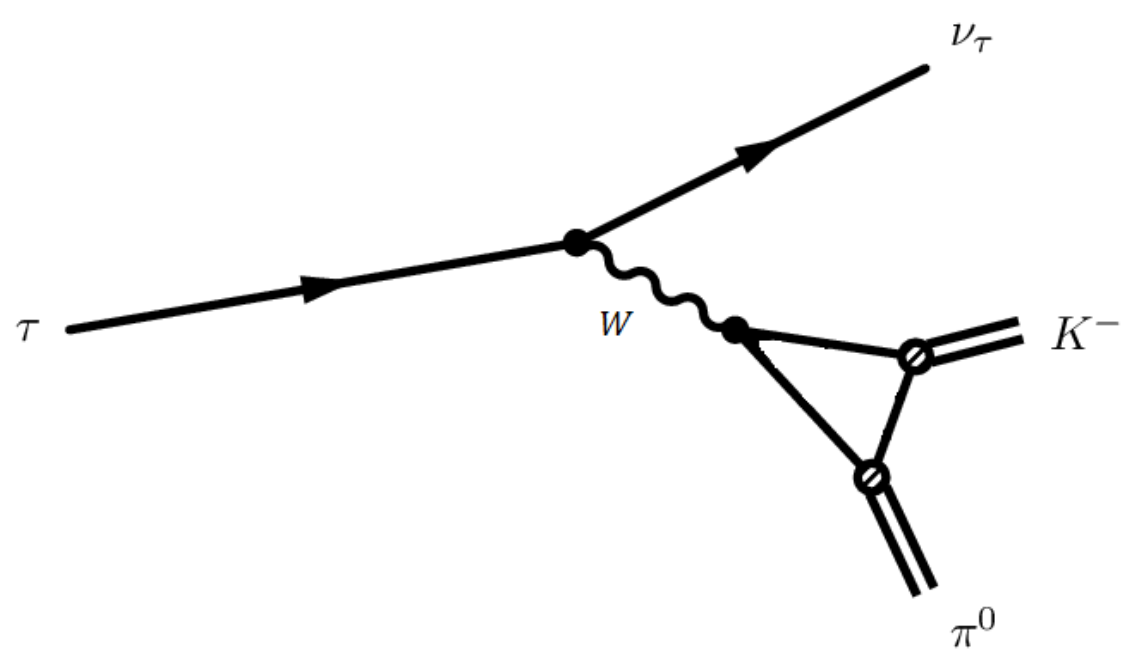}}
\caption{The decay $\tau \rightarrow K^{-}\pi^{0}\nu_{\tau}$ with intermediate $W$-boson (Contact diagram)}
\label{Contact}
\end{figure}

\begin{figure}[b]
\centerline{\includegraphics[scale = 0.7]{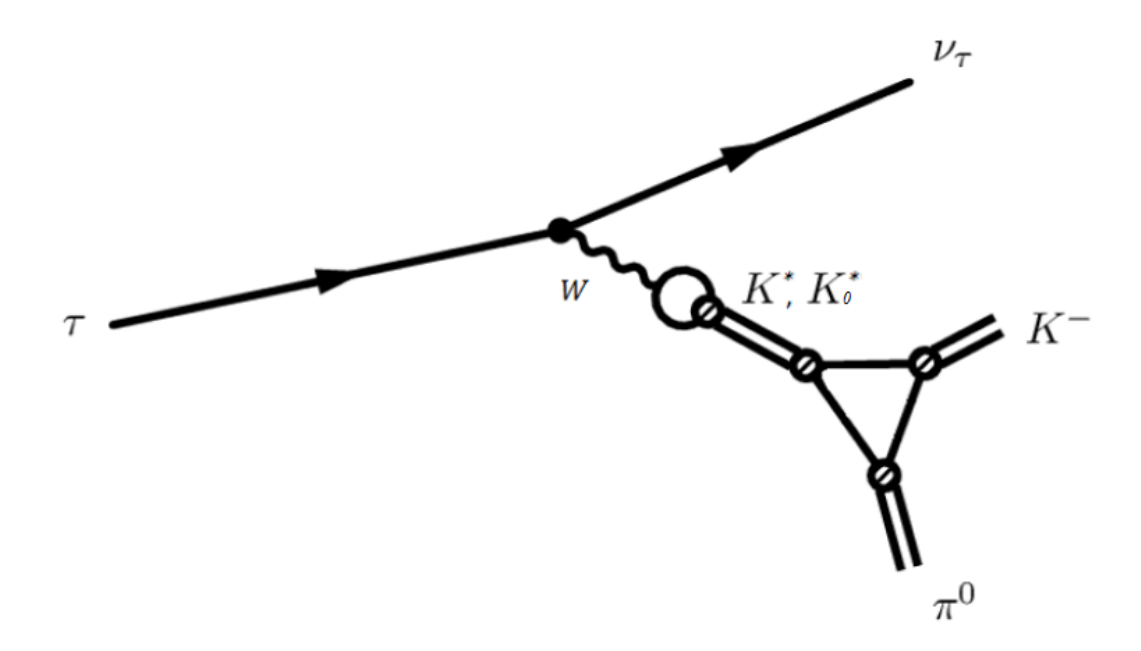}}
\caption{The decay $\tau \rightarrow K^{-}\pi^{0}\nu_{\tau}$ with intermediate vector $K^{*}(892)$ and scalar
$K_{0}^{*}(800)$ mesons}
\label{Intermediate}
\end{figure}

The diagrams of the process $\tau \rightarrow K^{-}\pi^{0}\nu_{\tau}$ considered in this paper
are shown in Fig.~\ref{Contact},\ref{Intermediate}


The amplitude of the process for the vector channel takes the form:
\begin{displaymath}
T_{V} = -\frac{i}{2} G_{F}V_{us} \sqrt{Z_{\pi}Z_{K}} \frac{g(m_{u}, m_{u})}{g(m_{u}, m_{s})} \left(\bar{\nu}_{\tau}\gamma^{\mu}\tau\right)
\end{displaymath}
\begin{equation}
\times \left\{g_{\mu\nu} +
\frac{\left[g_{\mu\nu}\left(q^{2} - \frac{3}{2}(m_{s} - m_{u})^{2}\right) - q_{\mu}q_{\nu}\right]}
{M_{K^{*}}^{2} - q^{2} - i\sqrt{q^{2}}\Gamma_{K^{*}}}\right\} (p_{K} - p_{\pi})^{\nu},
\end{equation}
where $G_{F} = 1.16637 \cdot 10^{-11}$ MeV$^{-2}$ is the Fermi constant, $V_{us} = 0.2252$ is the element of the Cabbibo-Kobayashi-Maskawa
matrix, $q = p_{K} + p_{\pi}$, $M_{K^{*}} = 892$ MeV and $\Gamma_{K^{*}} = 51$ MeV are the mass and the full
width of the vector meson \cite{Agashe:2014kda}.

The first term in the curly brackets corresponds to the diagram with the intermediate $W$-boson, the second term corresponds
to the diagram with the intermediate vector $K^{*}(892)$ meson.

The amplitude of the process for the scalar channel takes the form:
\begin{equation}
T_{S} = -\frac{i}{2} G_{F}V_{us} \sqrt{Z_{\pi}Z_{K}} \frac{g(m_{u}, m_{u})}{g(m_{u}, m_{s})} \left(\bar{\nu}_{\tau}\gamma^{\mu}\tau\right)
\frac{2m_{s}(m_{s} - m_{u})}{M_{K^{*}_{0}}^{2} - q^{2} - i\sqrt{q^{2}}\Gamma_{K^{*}_{0}}}q_{\mu},
\end{equation}
where $M_{K_{0}^{*}} = 682$ MeV, $\Gamma_{K_{0}^{*}} = 547$ MeV are the mass and full width of the scalar meson
(\cite{Agashe:2014kda}, p.949).

\section{Numerical estimations}
Following our calculations, the contribution of the diagrams with the vector channels to the branching of the process
$\tau \rightarrow K^{-}\pi^{0}\nu_{\tau}$  is equal to:
\begin{equation}
\textrm{Br}(\tau \rightarrow K^{-}\pi^{0}\nu_{\tau})_{V} = 4.14 \cdot 10^{-3}.
\end{equation}

The calculated contribution of the single diagram with the scalar meson is equal to:
\begin{equation}
\textrm{Br}(\tau \rightarrow K^{-}\pi^{0}\nu_{\tau})_{S} = 0.02 \cdot 10^{-3}.
\end{equation}

The calculated branching of the whole process is equal to:
\begin{equation}
\textrm{Br}(\tau \rightarrow K^{-}\pi^{0}\nu_{\tau})_{tot} = 4.13 \cdot 10^{-3}.
\end{equation}

The experimental value of this branching is equal to \cite{Agashe:2014kda}
\begin{equation}
\textrm{Br}(\tau \rightarrow K^{-}\pi^{0}\nu_{\tau})_{exp} = (4.29 \pm 0.15) \cdot 10^{-3}.
\end{equation}

\begin{figure}[h]
\centerline{\includegraphics[scale = 0.7]{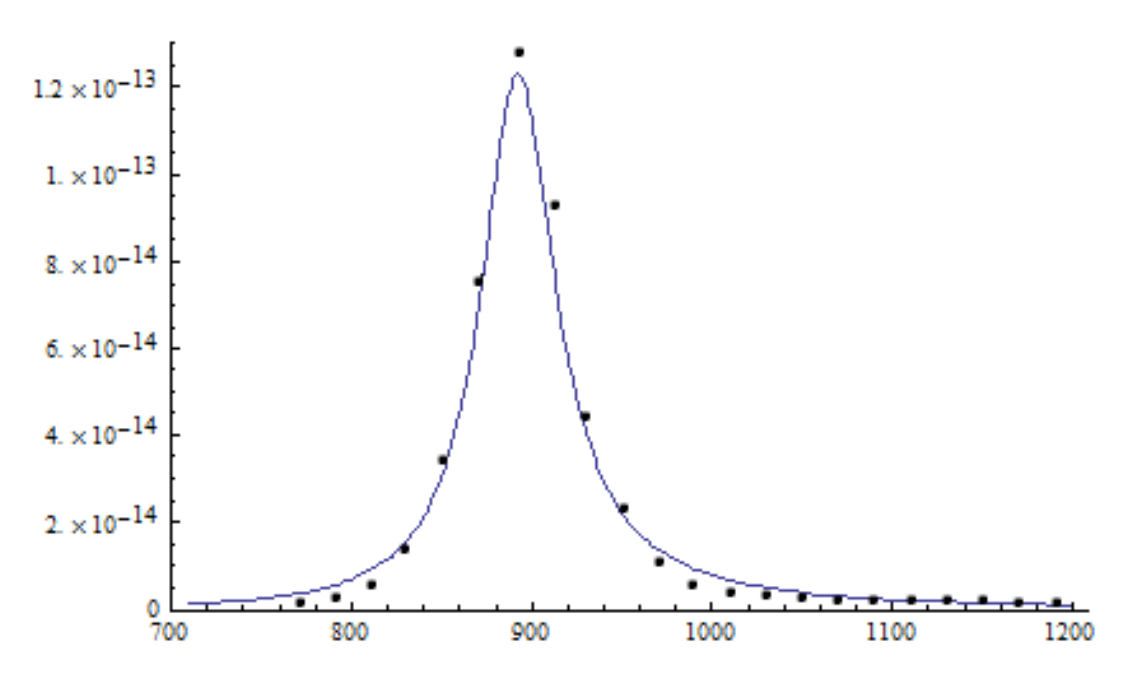}}
\caption{Differential width of the decay $\tau \rightarrow K^{-}\pi^{0}\nu_{\tau}$}
\label{Diff}
\end{figure}

The comparison of calculated and experimental differential width is shown in Fig.~\ref{Diff}. The solid line corresponds to our
theoretical differential width. The points correspond to the experimental values \cite{Aubert:2007jh}.

One can see that our results are in satisfactory agreement with the experimental data.

\section{Conclusion}
In the present work, the standard NJL model was used for description of the decay
$\tau \rightarrow K^{-}\pi^{0}\nu_{\tau}$. It was shown that the diagrams with intermediate
$W$-boson and intermediate vector $K^{*}(892)$ meson give the main contribution. Wherein the
dominant contribution is made by the second diagram with $K^{*}(892)$. Besides that, it is shown that the subprocess with the
scalar $K_{0}^{*}(800)$ meson changes insignificantly the result obtained from diagrams with the vector channel.
It was shown that our results of calculation of the width of the decay $\tau \rightarrow K^{-}\pi^{0}\nu_{\tau}$ in the framework
of the standard Nambu-Jona-Lasinio model are in a good agreement with experimental data on the branching and the differential
width of this process.

The NJL-like model was applied in \cite{Bednyakov:1992qv}. However,
method of the vector dominance was used there for description of the transition $W \rightarrow K^{*}$. Thus, the results of this work are
only in qualitative agreement with the experimental data. Nevertheless, the estimation of contribution of the scalar channel corresponds to
our results. In a number of other works, the models close to the models of Chiral Perturbation Theory
for research of $\tau \rightarrow K^{-}\pi^{0}\nu_{\tau}$ \cite{Finkemeier:1996dh, Jamin:2006tk, Jamin:2008qg,
Boito:2008fq, Boito:2010me} were applied. Our results are in qualitative agreement with their results.

In Appendix, it is shown that the contribution of the diagram with the intermediate radially excited ${K^{*}}'(1410)$
meson is small. For estimation of this contribution the extended NJL model was used.

\section*{Appendix: The contribution of the intermediate radially excited vector ${K^{*}}'(1410)$ meson}

For estimations of the contribution of the intermediate excited vector ${K^{*}}'(1410)$ meson to the process
$\tau \rightarrow K^{-}\pi^{0}\nu_{\tau}$ one can apply the extended NJL model
\cite{Volkov:2006vq, Volkov:1996br, Volkov:1996rc, Volkov:1997dd, Volkov:1999yi}.
The appropriate Lagrangian takes the form:
\begin{displaymath}
\Delta L_{int}(q,\bar{q},\pi^{0},K,K_{0}^{*},K^{*}) = \bar{q}\left[i\gamma^{5}(a_{\pi}\pi^{0} + b_{\pi}\pi^{'0})\lambda_{3}\right.
\end{displaymath}
\begin{equation}
\left. + i\gamma^{5}(a_{K}K^{\pm} + b_{K}K^{'\pm})\lambda_{\pm} +
\frac{1}{2}\gamma^{\mu}(a_{K^{*}}K^{*\pm}_{\mu} + b_{K^{*}}K^{*'\pm}_{\mu})\lambda_{\pm}\right]q,
\end{equation}
where
\begin{displaymath}
a_{a} = \frac{1}{\textrm{sin}\left(2\theta_{a}^{0}\right)}\left[g_{a}\textrm{sin}\left(\theta_{a} + \theta_{a}^{0}\right) +
g_{a}'f_{a}\left(k_{\perp}^{2}\right)\textrm{sin}\left(\theta_{a} - \theta_{a}^{0}\right)\right],
\end{displaymath}
\begin{equation}
b_{a} = \frac{-1}{\textrm{sin}\left(2\theta_{a}^{0}\right)}\left[g_{a}\textrm{cos}\left(\theta_{a} + \theta_{a}^{0}\right) +
g_{a}'f_{a}\left(k_{\perp}^{2}\right)\textrm{cos}\left(\theta_{a} - \theta_{a}^{0}\right)\right],
\end{equation}
$f_{a}\left(k_{\perp}^{2}\right) = 1 + d_{a}k_{\perp}^{2}$ is the form factor, $d_{a}$ is the slope parameter
($d_{us} = -1.75$GeV$^{-2}$), $\theta_{a}$ and $\theta_{a}^{0}$ are the mixing angles calculated using the method
presented in \cite{Volkov:1999yi}:
\begin{equation}
\begin{array}{ccc}
\theta_{\pi} = 59.48^{\circ}     & \theta_{K} = 70.68^{\circ}      & \theta_{K^{*}} = 87.49^{\circ}\\
\theta_{\pi}^{0} = 59.12^{\circ} & \theta_{K}^{0} = 55.77^{\circ}  & \theta_{K^{*}}^{0} = 59.92^{\circ}
\end{array}
\end{equation}

The fields in the excited states are marked with prime.

The coupling constants:
\begin{displaymath}
g_{\pi}' = \left(4I_{2}^{f^{2}}(m_{u}, m_{u})\right)^{-1/2}, \quad g_{K}' = \left(4I_{2}^{f^{2}}(m_{u}, m_{s})\right)^{-1/2},
\end{displaymath}
\begin{equation}
g_{K^{*}}' = \left(\frac{2}{3}I_{2}^{f^{2}}(m_{u}, m_{s})\right)^{-1/2},
\end{equation}
$g_{\pi}, g_{K}$ and $g_{K^{*}}$ are defined in (\ref{Constants}),
\begin{equation}
I_{2}^{f^{n}}(m_{1}, m_{2}) =
-i\frac{N_{c}}{(2\pi)^{4}}\int\frac{f^{n}(\vec{k}^{2})}{(m_{1}^{2} - k^2)(m_{2}^{2} - k^2)}\Theta(\Lambda_{3}^{2} - \vec{k}^2)
\mathrm{d}^{4}k,
\end{equation}
$\Lambda_{3} = 1.03$ GeV is the cut-off parameter.

The amplitude of the process $\tau \rightarrow K^{-}\pi^{0}\nu_{\tau}$ for the intermediate
excited vector ${K^{*}}'(1410)$ meson is
\begin{displaymath}
T_{exc} = -\frac{i}{2} G_{F}V_{us} \sqrt{Z_{\pi}Z_{K}} \frac{g(m_{u}, m_{u})}{g(m_{u}, m_{s})}  C_{1}C_{2} \left(\bar{\nu}_{\tau}\gamma^{\mu}\tau\right)
\end{displaymath}
\begin{equation}
\times \frac{g_{\mu\nu}\left[q^{2}-\frac{3}{2}(m_{s} - m_{u})^{2}\right] - q_{\mu}q_{\nu}}{M_{{K^{*}}'}^{2} - q^{2} - i\sqrt{q^{2}}\Gamma_{{K^{*}}'}}(p_{K} - p_{\pi})^{\nu},
\end{equation}
where $M_{{K^{*}}'} = 1410$ MeV, $\Gamma_{{K^{*}}'} = 232$ MeV are the mass and the full width of the excited ${K^{*}}'(1410)$ meson \cite{Agashe:2014kda},
\begin{displaymath}
C_{1} = \frac{-1}{\textrm{sin}\left(2\theta_{K^{*}}^{0}\right)}\left[\textrm{cos}\left(\theta_{K^{*}} + \theta_{K^{*}}^{0}\right) +
R_{V}\textrm{cos}\left(\theta_{K^{*}} - \theta_{K^{*}}^{0}\right)\right],
\end{displaymath}
\begin{displaymath}
C_{2} = \frac{-1}{\textrm{sin}\left(2\theta_{K^{*}}^{0}\right)\textrm{sin}\left(2\theta_{K}^{0}\right)}\left[
\textrm{cos}\left(\theta_{K^{*}} + \theta_{K^{*}}^{0}\right)\textrm{sin}\left(\theta_{K} + \theta_{K}^{0}\right)\right.
\end{displaymath}
\begin{displaymath}
+ \frac{1}{\sqrt{Z_{K}}}\textrm{cos}\left(\theta_{K^{*}} - \theta_{K^{*}}^{0}\right)\textrm{sin}\left(\theta_{K} - \theta_{K}^{0}\right) + R_{V}\textrm{cos}\left(\theta_{K^{*}} - \theta_{K^{*}}^{0}\right)\textrm{sin}\left(\theta_{K} + \theta_{K}^{0}\right)
\end{displaymath}
\begin{displaymath}
\left. + R_{S}\textrm{cos}\left(\theta_{K^{*}} + \theta_{K^{*}}^{0}\right)\textrm{sin}\left(\theta_{K} - \theta_{K}^{0}\right)\right],
\end{displaymath}
\begin{equation}
R_{S} = \frac{I_{2}^{f}(m_{u},m_{s})}{\sqrt{Z_{K}I_{2}(m_{u},m_{s})I_{2}^{f^{2}}(m_{u},m_{s})}}, \quad
R_{V} = \frac{I_{2}^{f}(m_{u},m_{s})}{\sqrt{I_{2}(m_{u},m_{s})I_{2}^{f^{2}}(m_{u},m_{s})}}.
\end{equation}

The contribution of this subprocess is:
\begin{equation}
\textrm{Br}(\tau \rightarrow K^{-}\pi^{0}\nu_{\tau})_{exc} = 0.035 \cdot 10^{-3}.
\end{equation}

\section*{Acknowledgments}
We are grateful to A. B. Arbuzov, S. B. Gerasimov, D. G. Kostunin, O. V. Teryaev and A. V. Zarubin for useful
discussions.

\end{document}